%% file: main.tex
\documentclass[10pt,letterpaper]{article}
\input{preamble}

\begin{document}

\hypersetup{pageanchor=false}
\begin{titlepage}
\thispagestyle{empty}
\begin{center}
\begingroup
\renewcommand{\thefootnote}{\fnsymbol{footnote}}
{\LARGE\bfseries \paperTitle\footnote{Submitted to RECOMB 2027.}\par}
\endgroup
\setcounter{footnote}{0}
\vspace{1cm}
{\large Ronen Nir \quad and \quad Gal Keinan\par}
\vfill
\begin{minipage}{0.94\textwidth}
\textbf{Abstract.} Biomarkers can guide diagnosis and treatment decisions, but candidates identified from molecular data require independent experimental validation.
Because validation can be costly, nomination methods should prioritize compact candidate sets with few false discoveries.
Correlations among omics measurements complicate this task, making it difficult to limit false discoveries while maintaining recovery power.
We present Select-then-Knock (STK), a multi-split method for biomarker nomination that combines elastic-net stability selection with approximate Model-X knockoff filtering.
In each repetition, STK screens candidates on one sample subset and tests them on a disjoint subset.
It then retains features selected in at least a prespecified fraction of repetitions.

We analyze this frequency-based aggregation to characterize when it suppresses false selections and when it reinforces persistent errors.
We derive deterministic bounds relating agreement across splits to the number of retained features, along with conditional bounds on expected false-selection counts under independent, identically distributed repetitions on a fixed dataset.

Across benchmarks, STK generally selected smaller feature sets with lower mean false discovery proportion (FDP) than Stabl, which calibrates stability selection using artificial control features.
Recovery power was comparable or lower.
A separate study isolating the effect of aggregation showed that combining selections across splits reduced mean FDP, sometimes at a cost in power.
In tests with no features associated with the outcome, STK selected no candidates, whereas several comparison methods frequently made false selections.
At 20,000 features, STK reduced peak memory by a factor of 7.1 and mean runtime by a factor of 1.6 relative to direct full-dimensional Model-X knockoff filtering.
These results support STK as a computationally efficient approach to compact biomarker nomination, although its final support has no proven false discovery rate (FDR) guarantee, and benchmarks with synthetic outcomes do not establish biological validity.
\end{minipage}
\vspace{0.5cm}
\begin{minipage}{0.94\textwidth}
\end{minipage}
\vfill
\end{center}
\end{titlepage}
\clearpage
\pagenumbering{arabic}
\hypersetup{pageanchor=true}

\section{Introduction}
\label{sec:introduction}

Biomarker studies often measure thousands of features, but only a compact set can feasibly proceed to assay development, perturbational testing, or other experimental follow-up.
Choosing that set is difficult when sample sizes are modest and features are strongly correlated: signatures with similar predictive performance can change substantially with the training sample \citep{eindor2005outcome}.
Predictive accuracy alone is therefore insufficient for identifying a consistent set of candidates.
Our objective is to nominate compact sets of associated features for further study, favoring candidates that recur across sample partitions.
These nominations require independent experimental validation; selection alone establishes neither biomarker validity nor causality \citep{poste2011biomarkers}.

Elastic-net regression encourages sparse models while accommodating correlated features \citep{zou2005regularization}.
Stability selection addresses sensitivity to sampling by retaining features that recur across subsamples \citep{meinshausen2010stability}.
Model-X knockoffs provide a complementary comparison: each measured feature is paired with a synthetic decoy, and selection depends on evidence favoring the original over its decoy \citep{barber2015controlling,candes2018panning}.
Applying knockoff filtering to the full feature set requires constructing these decoys and fitting them alongside the original features.
At transcriptome scale, this can impose substantial computational and memory costs.

Screening before knockoff filtering reduces the number of features that require decoys, but using the same outcome data for screening and testing can introduce selection bias \citep{ambroise2002selection,barber2019highdimensional}.
Features are first retained because they appear associated with the outcome, then assessed using the same evidence.
Using disjoint subsets of observations for screening and testing avoids this reuse \citep{wasserman2009highdimensional,barber2019highdimensional}.
A single split, however, leaves fewer observations for each stage and can make the selected set sensitive to the partition.
This motivates combining selections across partitions for compact nomination and examining what such aggregation does---and does not---imply about false discoveries.

We introduce \textsc{Select-then-Knock} (STK), a multi-split pipeline for compact biomarker nomination.\footnote{Source code: \sourceCodeURL.}
For each random partition, STK tunes an elastic net and runs stability selection on the screening subset.
On the disjoint testing subset, it constructs approximate Model-X knockoffs only for the resulting candidates and applies the knockoff filter.
STK retains features selected in at least a prespecified fraction of partitions.
Unlike derandomized knockoffs, which repeat knockoff draws on fixed data and a fixed set of tested features \citep{ren2023derandomizing}, STK repeats both screening and held-out filtering over new partitions.
The candidate set, testing observations, and knockoff draw can therefore all change between repetitions.

We provide a theoretical analysis of the frequency aggregation rule.
We formally define the repeated-selection process and prove deterministic bounds relating overlap among split selections to the number of retained features, without assuming independent splits.
We then analyze an idealized model in which repetitions on a fixed dataset are independent and follow the same distribution.
Under this model, we derive feature-retention probabilities and bounds on the expected number of false selections for a finite number of repetitions.
These results explain how aggregation can suppress false selections that recur infrequently but reinforce those that recur consistently.
Finally, a counterexample for a generic selector proves that per-split false discovery rate (FDR) control alone does not guarantee FDR control after aggregation (Section~\ref{sec:theory}; Appendix~\ref{app:theory}).
We do not claim FDR or per-family error rate (PFER) control for STK's final support.

We evaluate support recovery, false selections, and computational cost in synthetic Gaussian and semi-synthetic DepMap benchmarks, global-null experiments, and a 20,000-feature scaling study.
Matched split analyses isolate the effect of aggregation.
Our principal empirical comparator is Stabl, which calibrates stability selection using artificial control features \citep{hedou2024stabl}.
PC-Knockoff provides a screened-knockoff comparison: it screens features on one sample subset and applies knockoff filtering on another \citep{liu2022modelfree}.
We evaluate both single-split PC-Knockoff and a repeated-PC adaptation using the same frequency rule as STK.

\section{Related Work}
\label{sec:related-work}

STK is closest to methods that reduce the feature family before knockoff
inference. Barber and Cand\`es use sample splitting to screen features before
fixed-design knockoff filtering in high-dimensional linear models
\citep{barber2019highdimensional}. PC-Knockoff similarly screens on one row
subset and applies knockoffs to the retained features on another
\citep{liu2022modelfree}. STK shares this separation but uses elastic-net
stability selection over all features, applies approximate Model-X knockoff
filtering to the resulting candidates, and repeats the complete procedure over
row partitions. Its contribution is therefore this pipeline composition and
its empirical behavior and aggregation analysis, not the screen-then-knockoff pattern itself.

Two related methods clarify the role of aggregation and artificial controls.
Derandomized knockoffs repeat knockoff generation on a fixed dataset and retain
features by their selection frequency \citep{ren2023derandomizing}; STK instead
redraws the row partition, so both the candidate family and held-out test change
between repetitions. Stabl targets sparse omic biomarkers by combining
stability selection with knockoff or permuted artificial features and using
their selection frequencies to calibrate a reliability threshold
\citep{hedou2024stabl}.
In contrast, STK uses stability selection to form candidates before separate, held-out knockoff filtering.
STK forms its final support by frequency thresholding, without a proved FDR or PFER guarantee.

Our aggregation analysis draws on established stability and counting ideas
\citep{meinshausen2010stability,shah2013variable,ren2023derandomizing}.
The recurrence ratio is related to the stability index of
\citet{nogueira2018stability}; the bounds use agreement to describe retention,
not to certify biological relevance.

\section{Method}
\label{sec:method}

\subsection{Pipeline overview}
\label{sec:method-overview}

\begin{figure}[t]
  \centering
  \includegraphics[width=\textwidth]{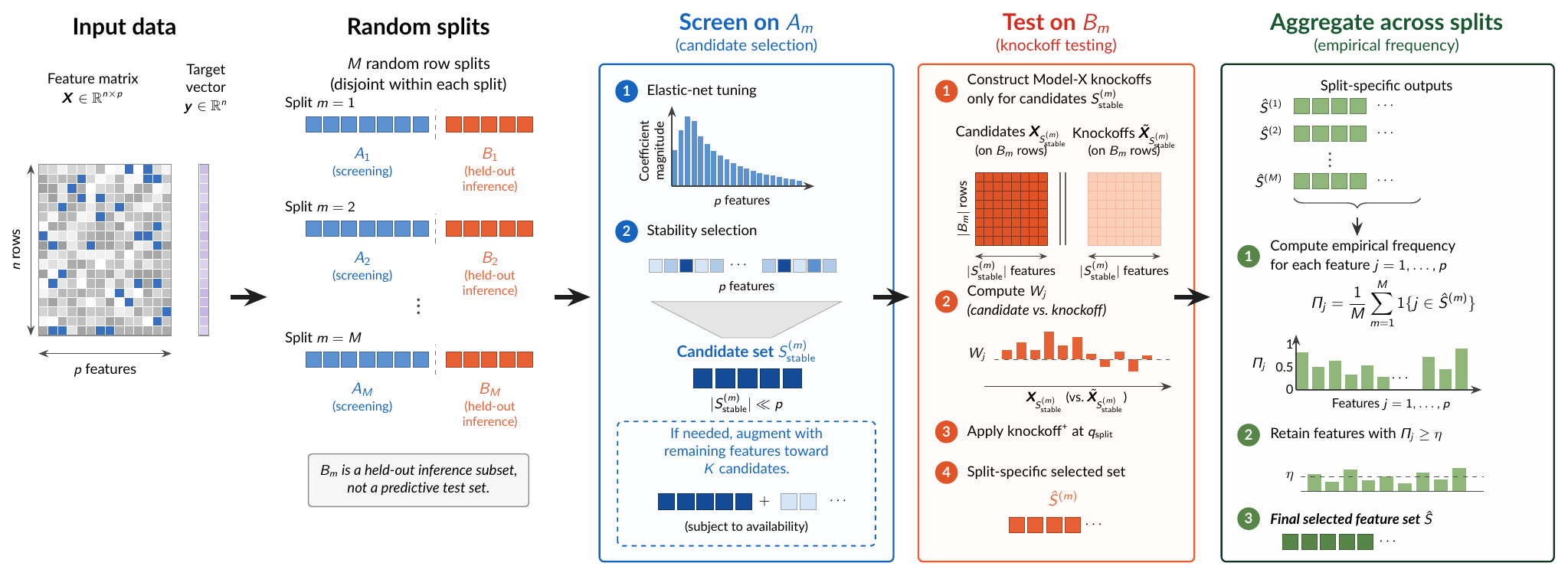}
  \caption{\textbf{Multi-split Select-then-Knock.}
  STK repeats screening and held-out testing over $M$ random row partitions.
  In partition $m$, elastic-net tuning and stability selection run on screening rows $A_m$.
  If fewer than $K$ features pass the stability threshold, the highest-scoring remaining features are added to form the candidate set $\Sstable^{(m)}$.
  On the disjoint rows $B_m$, STK constructs approximate Model-X knockoffs only for these candidates and applies the knockoff filter at level $q_{\mathrm{split}}$, producing $\hat{S}^{(m)}$.
  The final support $\hat{S}$ contains features selected in at least an $\eta$ fraction of the partitions.}
  \label{fig:pipeline}
\end{figure}

Given a feature matrix $\mX \in \R^{n \times p}$ and target vector $\vy \in \R^n$, STK seeks a small support $\hat{S}\subseteq\{1,\dots,p\}$ of associated features rather than a prediction-optimal model.
STK draws $M$ random partitions of the $n$ rows (samples).
Partition $m$ splits the rows into two disjoint sets: a screening set $A_m$ containing a fraction $\gamma$ of the rows, and a testing set $B_m$ containing the rest.
The screening stage tunes an elastic net and runs stability selection using only $A_m$.
The testing stage constructs approximate Model-X knockoffs for the screened candidates and applies the knockoff filter using only $B_m$.
Screening and testing use disjoint rows within each partition, although rows may recur across partitions.
Finally, STK aggregates the selected features across partitions.
Figure~\ref{fig:pipeline} describes this process end to end.

\subsection{Screening stage: stability selection}
\label{sec:screening}

On screening rows $A_m$, STK first selects the elastic-net regularization strength $\lambdaopt$ and $\ell_1/\ell_2$ mixing $\alphaopt$ by cross-validation.
It then reuses these values for stability selection \citep{meinshausen2010stability}, repeatedly fitting the elastic net to subsamples of $A_m$.
The stability score $\piphat_j$ is the fraction of successful subsample fits in which feature $j$ has a nonzero coefficient.

Let $\bar q_{\mathrm{stab}}$ be the mean number of selected features across successful fits, $p_m$ the number of nonconstant screening features in split $m$, and $V_{\mathrm{stab}}$ a screening calibration parameter.
STK uses the threshold
\begin{equation}
  \pi_{\mathrm{thr}}=1/2+\frac{\bar q_{\mathrm{stab}}^{\,2}}
  {2p_mV_{\mathrm{stab}}},
\end{equation}
obtained by inverting the Meinshausen--B\"uhlmann bound and clipped to $[0.51,0.99]$.
Features with $\piphat_j\geq\pi_{\mathrm{thr}}$ form the initial stable set.
If it contains fewer than the configured candidate floor $K$, STK adds the next highest-scoring features until it reaches $K$ or exhausts the available features.
We denote the resulting candidate set for split $m$ by $\Sstable^{(m)}$.
Only this candidate set, rather than all $p$ features, passes to the computationally intensive testing stage.
The screening threshold does not guarantee PFER (the expected number of false selections) for the augmented candidate set or error control for the final support.
$K$ is a floor, not a cap.

\subsection{Testing stage: knockoff filtering}
\label{sec:testing}

On the held-out rows $B_m$, STK applies a Model-X knockoff filter \citep{candes2018panning,barber2015controlling} only to the columns indexed by $\Sstable^{(m)}$.
It constructs second-order Gaussian decoy features from the held-out candidate matrix and fits a cross-validated lasso to the original and decoy features.
Because the feature distribution is estimated, these knockoffs are approximate.
Let $\beta_j$ and $\tilde\beta_j$ denote the fitted coefficients of candidate $j$ and its knockoff in the joint lasso fit.
The statistic $W_j=|\beta_j|-|\tilde\beta_j|$ is positive when the original feature has the larger absolute coefficient.
Given a knockoff level $q_{\mathrm{split}}$ and numerator offset $a$, the filter chooses the smallest qualifying observed absolute statistic.
With $\mathcal W=\{|W_j|:|W_j|>0\}$, set
\begin{equation}
  T = \min\!\Big\{ t\in\mathcal W \;:\;
  \frac{a + \#\{j:W_j \le -t\}}
  {\max\!\big(1,\#\{j:W_j \ge t\}\big)}
  \le q_{\mathrm{split}} \Big\},
\end{equation}
with $T=\infty$ if no statistic qualifies, and return $\hat S^{(m)}=\{j:W_j\geq T\}$.
The reported experiments use Knockoff\textsuperscript{+} with $a=1$; a nonempty result then requires at least $\lceil1/q_{\mathrm{split}}\rceil$ selected features.
We use $q_{\mathrm{split}}=0.5$ to allow small nonempty selections: a split must select at least two features or none, whereas $q_{\mathrm{split}}=0.1$ requires at least ten.
The testing level $q_{\mathrm{split}}$ is separate from the quantities used to set the screening threshold, $\bar q_{\mathrm{stab}}$ and $V_{\mathrm{stab}}$.

\subsection{Multi-split aggregation}
\label{sec:multisplit}

Multi-split STK repeats the screening and testing stages over $M$ row partitions.
Each partition produces its own candidate set, held-out testing rows, knockoff draw, and selected set $\hat S^{(m)}$.
For feature $j$, define its selection frequency as
\begin{equation}
  \Pi_j=\frac{1}{M}\sum_{m=1}^{M}\mathbf{1}\{j\in\hat S^{(m)}\}.
\end{equation}
STK retains features selected in at least a fraction $\eta$ of the partitions, $\hat S=\{j:\Pi_j\geq\eta\}$.
The evaluated multi-split configurations use $M=25$ and $\eta=1/2$: a feature must therefore be selected in at least 13 of the 25 partitions to enter the final support.

Table~\ref{tab:operating-points} summarizes each parameter's role; the complete implementation and evaluated configurations are provided with the code and reproduction materials.

The effect of this frequency rule depends on how selections recur across partitions.
Section~\ref{sec:theory} relates agreement across splits to the selections retained by aggregation, then examines conditions under which aggregation suppresses or amplifies false selections.

\Needspace{14\baselineskip}
\input{theory}

\section{Experimental Evaluation}
\label{sec:evaluation}

\subsection{Benchmark designs}

We evaluated STK's support recovery, false selections, and computational cost using the designs in Table~\ref{tab:benchmarks}.
The synthetic Gaussian and semi-synthetic DepMap benchmarks assessed the precision--power trade-off.
Global-null experiments tested abstention, and a scaling study measured runtime and memory use.
To isolate aggregation, we compared final-support FDP and power with their means across the same 25 split outputs used to construct that support (Section~\ref{sec:matched}).
We compared STK with PC-Knockoff on a separate set of synthetic Gaussian datasets (Appendix~\ref{app:pc}).

\begin{table}[H]
  \centering
  \scriptsize
  \setlength{\tabcolsep}{3pt}
  \caption{Experimental benchmark designs.}
  \label{tab:benchmarks}
  \begin{tabular}{@{}p{0.22\columnwidth}p{0.70\columnwidth}@{}}
    \toprule
    Setting & Design and repetitions \\
    \midrule
    Synthetic Gaussian & Block-correlated Gaussian features with
    $n=800$, $p=5000$, block size 20, $\rho=0.5$, and signal-to-noise ratio
    0.5; true support size $k\in\{1,2,3,5,10,15,20\}$, with support features
    in distinct blocks; ten seeds per $k$. \\
    DepMap & DepMap 26Q1 protein-coding RNA-seq
    \citep{depmap2026q1}; 800 cell lines and 19,201 retained features after
    removing 14 zero-variance features from 19,215 inputs; synthetic linear
    outcomes with true support size $k\in\{3,10\}$, signal-to-noise ratio 2,
    and pairwise absolute correlation at most 0.5 among support features;
    ten datasets per $k$. \\
    Matched aggregation & Recorded outputs from 25 splits per dataset in the
    synthetic Gaussian benchmark design; ten datasets at each of the seven support sizes
    and 30 global-null datasets at $p=5000$, $\rho=0.5$. \\
    Screened knockoffs & Synthetic Gaussian datasets at $k\in\{1,3,10,20\}$
    (10 seeds per $k$), plus 30 global-null datasets (seeds 10000--10029);
    the same feature matrices and outcomes for STK and PC-Knockoff, with single-split PC and
    repeated PC over 25 splits. \\
    Gaussian global null & Block-correlated Gaussian features with $n=800$ and targets
    drawn independently from $N(0,1)$; $\rho\in\{0,0.3,0.5,0.8\}$ at $p=5000$
    and $p\in\{500,2000,5000\}$ at $\rho=0.5$; 30 seeds per setting. \\
    Scaling & Block-correlated Gaussian features with $n=800$,
    $k=3$, $\rho=0.8$, and signal-to-noise ratio 0.5;
    $p\in\{500,2000,5000,20000\}$ over ten seeds; 16 CPUs and 64~GB for both
    multi-split STK and direct full-$p$ Model-X. \\
    \bottomrule
  \end{tabular}
\end{table}

The true support $S^\star$ consists of the features with nonzero coefficients in the benchmark's outcome-generating model.
DepMap supplies realistic transcriptomic covariance, but its synthetic outcomes do not validate biological relevance or drug response.

\subsection{Comparators and operating points}
\label{sec:comparators}

Knockoff-based Stabl \citep{hedou2024stabl} is our principal empirical comparator, alongside LassoCV \citep{tibshirani1996regression} and the same-sample pooled STK reference defined below.
We evaluated PC-Knockoff \citep{liu2022modelfree} as a screened-knockoff comparator.
For the main PC comparison, we used 200 screening rows, 600 testing rows, and $d=50$ candidates at three prespecified thresholds, $\alpha\in\{0.1,0.2,0.5\}$.
Repeated PC aggregates 25 split outputs by the same frequency rule as STK; alternative allocations are detailed in Appendix~\ref{app:pc}.
Direct Model-X filtering over all $p$ features provides the computational-scaling reference.

Table~\ref{tab:operating-points} contrasts the two STK configurations.
Headline STK is our primary configuration for sparse-support recovery.
The secondary configuration increases the candidate floor and screening subset to test whether these changes improve denser-support recovery and to evaluate scaling with a larger candidate set.
We report it as a sensitivity analysis rather than as a replacement for the headline configuration.

\begin{table}[H]
  \centering
  \footnotesize
  \setlength{\tabcolsep}{3pt}
  \caption{Evaluated multi-split configurations. Values are shown as
  headline/secondary.}
  \label{tab:operating-points}
  \begin{tabular}{@{}lcp{0.46\columnwidth}@{}}
    \toprule
    Parameter & H/S & Role \\
    \midrule
    $M$ & 25/25 & Number of row partitions \\
    $\gamma$ & 0.50/0.70 & Screening-row fraction \\
    $K$ & 10/20 & Minimum candidate count \\
    $q_{\mathrm{split}}$ & 0.50/0.50 & Per-split knockoff level \\
    $\eta$ & 0.50/0.50 & Cross-split frequency threshold \\
    $a$ & 1/1 & Knockoff\textsuperscript{+} offset \\
    \bottomrule
  \end{tabular}
\end{table}

For comparison, we evaluate \emph{same-sample pooled STK} (hereafter pooled STK) as a reference configuration.
It uses the same observations to select and test candidates, without aggregating results across splits.
This variant is computationally convenient, but it double-dips: its tested candidate set depends on the same target values used for testing, so it does not inherit the standard knockoff guarantee.
Its comparison with multi-split STK is descriptive rather than a clean architectural ablation: the evaluated implementations differ in sample use, aggregation, per-filter level, and fitting budget, and the secondary configuration also uses a larger candidate floor.
The DepMap comparison with pooled STK is reported in Appendix~\ref{app:depmap-pooled}.
The scalability benchmark used the secondary configuration.

We report FDP, power, exact-support recovery frequency, and selected support size, computed in original feature coordinates.
FDP is the proportion of selected features outside the true support, defined as zero for an empty selection; power is the proportion of the true support recovered.

Mean FDP and power average dataset-level values; error bars show standard errors across datasets.
Paired comparisons use descriptive, unadjusted 95\% $t$ intervals calculated from dataset-level differences, not overlapping splits; an interval containing zero does not establish equivalence.
Mean FDP is an empirical summary, not an error-control guarantee.

\subsection{Gaussian precision--power trade-off}

\begin{figure}[t]
  \centering
  \includegraphics[width=\textwidth]{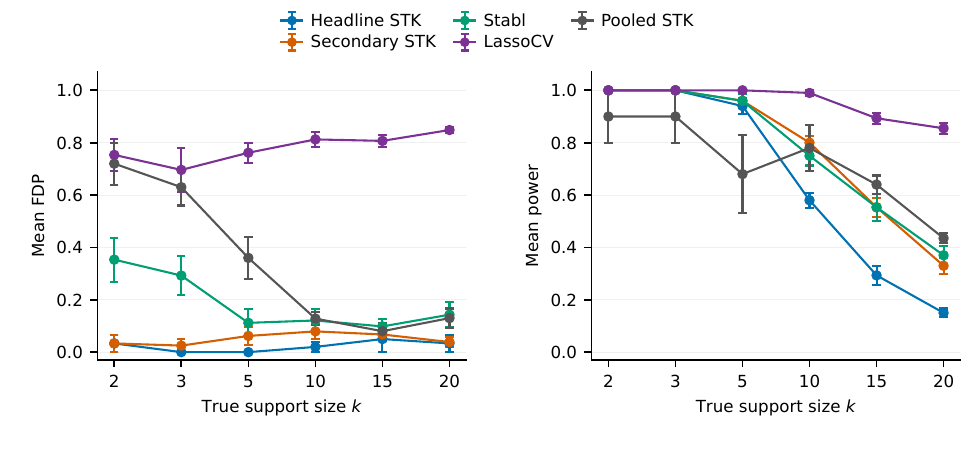}
  \caption{\textbf{Support recovery across true support sizes.}
  Mean FDP and power over ten synthetic Gaussian datasets per support size $k\geq2$; error bars show standard errors.
  Headline STK favors sparse precision, while secondary STK improves dense-support power.}
  \label{fig:density}
\end{figure}

We first tested how support density affects recovery in the synthetic Gaussian benchmark, where the correlation structure and true support are known (Figure~\ref{fig:density}).
At $k=3$, headline STK achieved power 1.00 with FDP 0 and recovered the exact support in all ten datasets.
Stabl achieved the same power, but with mean FDP 0.292 and exact recovery in 3 of 10 datasets.

As support grew, headline STK remained precise but lost power: at $k=10$, 15, and 20, its power was 0.58, 0.29, and 0.15, with mean FDP at most 0.05.
The secondary configuration recovered part of this loss.
At $k=10$, its power was 0.80 at FDP 0.079, compared with 0.75 at FDP 0.121 for Stabl.
At $k=15$, both had power 0.55, with FDP 0.067 and 0.098, respectively.
At $k=20$, secondary STK had power 0.33 at FDP 0.039, compared with Stabl's power 0.37 at FDP 0.143.
The larger candidate floor and screening subset therefore improved dense-support recovery in this benchmark, but produced a precision--power trade-off rather than uniform dominance.

On separate Gaussian datasets at $k=3,10,20$, both STK configurations had higher mean power and lower mean FDP than PC-Knockoff averaged over individual splits at the main PC setting with $\alpha=0.5$.
PC aggregation reduced FDP but also power at this setting, while larger screening allocations narrowed the power gap with headline STK (Appendix~\ref{app:pc}).

\subsection{Support recovery under transcriptomic covariance}

\begin{figure}[t]
  \centering
  \includegraphics[width=\textwidth]{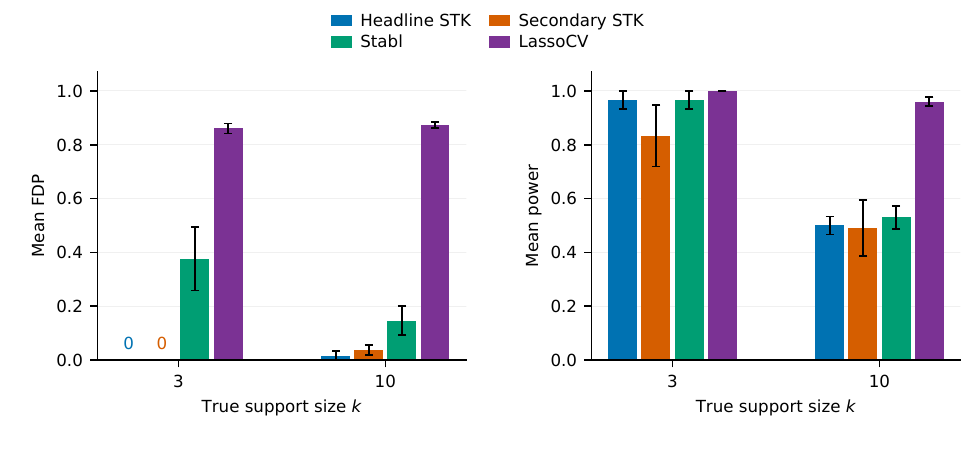}
  \caption{\textbf{Support recovery under transcriptomic covariance.}
  Mean FDP and power over ten semi-synthetic DepMap datasets per support size; error bars show standard errors.
  Headline STK had lower mean FDP than Stabl at both support sizes, with the same mean power at $k=3$ and slightly lower mean power at $k=10$.}
  \label{fig:depmap}
\end{figure}

We next evaluated support recovery using real DepMap expression measurements with synthetic linear outcomes (Figure~\ref{fig:depmap}).
At $k=3$, headline STK and Stabl both achieved mean power 0.967.
Headline STK selected 2.9 features on average with no observed false discoveries, compared with 8.0 features and mean FDP 0.377 for Stabl.
It recovered the true support exactly in 9 of 10 datasets, versus 4 of 10 for Stabl.
LassoCV achieved mean power 1.00 while selecting 24.7 features on average, with mean FDP 0.860.

At $k=10$, headline STK and Stabl achieved mean power 0.50 and 0.53, respectively, with mean FDP 0.017 and 0.147.
Across both support sizes, headline STK therefore had lower mean FDP than Stabl with nearly the same mean power.

The secondary STK configuration did not improve recovery on DepMap.
At $k=10$, its mean power was 0.49 and mean FDP was 0.037; at $k=3$, its mean power was 0.833, compared with 0.967 for headline STK.
The secondary configuration's power benefit at denser supports in the synthetic Gaussian benchmark therefore did not carry over to these DepMap settings.

\subsection{Effect of aggregation on FDP and power}
\label{sec:matched}

To measure the effect of aggregation, we compared the final support's FDP and power with the mean FDP and power across the 25 split outputs used to form it.
This separate matched experiment used the synthetic Gaussian benchmark design, with ten datasets per support size.
Aggregation lowered mean FDP in all 12 combinations of configuration and support size (Table~\ref{tab:matched-compact}).
The single-split value is the mean over the same 25 split outputs used for the aggregate, so this matched analysis does not compare independently run ensembles or equal compute budgets.

These precision gains sometimes came at a cost in power.
For the headline configuration, aggregate power fell at $k=15$ and 20; at $k=20$, it fell from 0.187 to 0.150.
Thus recurrence can remove selections that do not repeat, including relevant features.
These are descriptive results for the evaluated Gaussian design, not an error-control guarantee.

\begin{table}[htbp]
  \centering
  \caption{Matched aggregate versus mean individual-split STK performance.
   Differences are aggregate minus mean split; brackets are descriptive, unadjusted 95\% $t$ intervals across ten datasets per support size.
   Rows show $k\in\{3,10,20\}$; additional paired results for $k=2$, 5, and 15 are in Appendix~\ref{app:matched}.}
  \label{tab:matched-compact}
  \input{tables/matched_compact}
\end{table}

\subsection{False selections under the global null}

We next tested whether each method abstained when no feature was associated with the target.
Under the global null, every selection is false; we estimate the probability of selecting any feature, $P(|\hat S|>0)$, and report Wilson 95\% confidence intervals.
The Gaussian global-null benchmark varied correlation or feature dimension across six settings with 30 datasets each (Table~\ref{tab:benchmarks}).
Both multi-split STK configurations returned an empty support in all 180 datasets (Supplementary Figure~\ref{fig:null}).
By contrast, LassoCV selected at least one feature in 113 of 180 datasets (62.8\%), Stabl in 158 of 180 (87.8\%), and pooled STK in 125 of 180 (69.4\%).

In the separate matched global-null analysis ($p=5000$, $\rho=0.5$), the fraction of splits selecting any feature averaged 6.7\% for headline STK and 8.1\% for secondary STK across 30 datasets.
After aggregation, both configurations returned an empty support in all 30 datasets (Wilson 95\% interval $[0,0.114]$ for each configuration).
On the separate PC-comparison datasets, repeated PC selected features in at most 1 of 30 null datasets per evaluated setting (Appendix~\ref{app:pc}); abstention under the global null is therefore not unique to STK.
These analyses are not pooled and provide empirical evidence under the evaluated null designs, not a formal error-control guarantee.

\subsection{Computational scaling at transcriptome scale}

We compared secondary STK with direct full-dimensional Model-X filtering as the number of features increased from 500 to 20,000 (Supplementary Figure~\ref{fig:tractability}).
We used the secondary configuration to assess scaling with the larger candidate floor, $K=20$.
At $p=20{,}000$, secondary STK passed 20 candidates per split to knockoff construction, a 1,000-fold reduction in the tested family.
Peak resident memory, reported as the maximum across ten datasets, was 2.67~GiB for secondary STK and 18.84~GiB for direct Model-X, a 7.1-fold reduction.
Mean end-to-end runtime over the same ten datasets was 1,046 seconds for secondary STK and 1,641 seconds for direct Model-X, making STK 1.6-fold faster in this environment despite repeating 25 splits.

At this dimension, secondary STK achieved mean power 0.90 at mean FDP 0.058 and recovered the exact support in 6 of 10 datasets.
Direct Model-X completed successfully on all ten datasets without fallback but selected no features.
STK therefore reduced computational cost while retaining sparse-support recovery in this setting.

\section{Discussion and Limitations}
\label{sec:discussion}

STK is designed to nominate compact biomarker candidate sets when experimental follow-up is costly.
Across the synthetic Gaussian and semi-synthetic DepMap benchmarks, it generally selected smaller supports with lower mean FDP than Stabl, with comparable recovery power in some settings and lower power in others.
These results support compact nomination rather than exhaustive support recovery.
In the scaling study, secondary STK also ran faster and used less memory than direct full-dimensional Model-X filtering, despite repeating screening and testing across 25 partitions.

The theoretical analysis makes the role of recurrence explicit in two complementary ways.
First, the deterministic bounds relate recorded overlap across splits to both the fraction of selection occurrences retained and the number of distinct features in the aggregate (Proposition~\ref{prop:recurrence}).
They require no independence assumption and therefore apply even when split selections are dependent.
Within a feature set $G$, recurrence $\rho_G>\eta$ guarantees that at least one feature survives aggregation, while the upper bounds quantify how limited overlap restricts retention.

Second, the conditional IID analysis describes repeated randomization on a fixed dataset.
It gives exact binomial retention probabilities and a finite-repetition bound comparing the expected false-selection count after aggregation with that of one repetition (Corollary~\ref{cor:null-count}).
The bound depends on the largest conditional selection probability among null features: when the resulting multiplier is below one, it establishes a reduction in the expected false-selection count.
At the evaluated majority threshold, persistent false selections can instead be reinforced, while signals selected too rarely can be lost.
More repetitions therefore do not, by themselves, ensure better recovery.

The matched Gaussian results illustrate this retention trade-off: aggregation lowered mean FDP, but sometimes reduced power.
Repeated PC-Knockoff also suppressed null selections, so this benefit is not unique to STK.
The PC comparisons do not isolate the contribution of stability-based screening from the other differences between the pipelines.

Singleton supports ($k=1$) were evaluated in the Gaussian benchmarks but are reported only here.
In the synthetic Gaussian benchmark, headline STK selected 0.2 features on average and recovered the single true feature in 2 of 10 datasets, whereas Stabl recovered it in all ten at mean FDP 0.81.
Singleton recovery faces a structural obstacle: the split-level knockoff+ filter cannot retain the lone signal without also selecting another feature.
More generally, STK is not appropriate when a nonempty candidate set is required.

STK has no proved FDR or PFER guarantee for its final support.
Estimated Gaussian knockoffs may depart from Model-X exchangeability, and screening can change the conditional hypotheses being tested (Appendix~\ref{app:theory}).

DepMap supplied realistic transcriptomic covariance, but its outcomes were synthetic and linear, and predictive performance was not evaluated.
The experiments covered limited support sizes and seeds, and computational measurements are environment-specific.
The 25-repetition budget was a practical choice, not established by a convergence study.
Future studies should test broader assays and outcome models, assess sensitivity to the repetition budget, and develop formally controlled aggregation.
Nominated candidates still require independent experimental and domain-expert validation.

\Needspace{13\baselineskip}
\section{Conclusion}
\label{sec:conclusion}

We presented STK, a multi-split method combining stability-based screening, held-out knockoff filtering, and frequency aggregation for compact biomarker nomination.
Across the synthetic Gaussian and semi-synthetic DepMap benchmarks, STK generally returned smaller supports with lower mean FDP than Stabl, sometimes at a cost in recovery power.
Both multi-split configurations abstained under the evaluated Gaussian global-null settings.
In a separate transcriptome-scale study, secondary STK reduced runtime and memory use relative to direct full-dimensional Model-X filtering.

Our theory provides deterministic bounds on retained selections and conditional bounds on expected false-selection counts, clarifying how aggregation can suppress infrequent errors while reinforcing persistent ones.
STK has no proved FDR or PFER guarantee for its final support.
Together, these results support STK as a tractable approach to nominating compact candidate sets for independent experimental follow-up.

\clearpage
\bibliographystyle{plainnat}
\bibliography{references}

\clearpage
\appendix
\input{appendix}

\end{document}

%% file: preamble.tex
\usepackage[letterpaper,margin=1in]{geometry}
\usepackage[T1]{fontenc}
\usepackage[utf8]{inputenc}
\usepackage{lmodern}
\usepackage{microtype}
\usepackage{amsmath,amssymb,amsthm}
\usepackage{graphicx}
\usepackage{booktabs}
\usepackage{float}
\usepackage{needspace}
\usepackage[section]{placeins}
\usepackage[ruled,vlined]{algorithm2e}
\usepackage{caption}
\usepackage[numbers,sort&compress]{natbib}
\usepackage[hidelinks]{hyperref}
\renewcommand{\footnotesize}{\normalsize}
\renewcommand{\scriptsize}{\normalsize}

\newtheorem{proposition}{Proposition}
\newtheorem{corollary}[proposition]{Corollary}
\newcommand{\paperTitle}{Select-then-Knock: Multi-Split Feature Selection for Compact Biomarker Nomination}
\newcommand{\sourceCodeURL}{\url{https://github.com/GSK-AI/select-then-knock}}
\newcommand{\mX}{\boldsymbol{X}}
\newcommand{\vy}{\boldsymbol{y}}
\newcommand{\R}{\mathbb{R}}
\newcommand{\piphat}{\hat{\pi}}
\newcommand{\Sstable}{S_{\text{stable}}}

\newcommand{\alphaopt}{\alpha^{\star}}
\newcommand{\lambdaopt}{\lambda^{\star}}
\hypersetup{pdftitle={\paperTitle},pdfauthor={Ronen Nir and Gal Keinan}}

%% file: theory.tex
\section{Behavior of frequency aggregation}
\label{sec:theory}

Frequency aggregation retains features that recur across splits and discards those selected less often.
This can remove false selections, but it can also discard signals or retain persistent errors.
We analyze this trade-off in two steps: first, by showing how overlap between split selections constrains how many features survive aggregation, and then by examining how repeated randomization changes the expected number of false selections.
The first analysis is deterministic and requires no independence assumption; the second uses an idealized conditional model.

\paragraph{Retention among recorded selections.}
Let $N_j=\sum_{m=1}^M\mathbf{1}\{j\in\hat S^{(m)}\}$ count the splits selecting feature $j$, and let $r=\lceil M\eta\rceil$ be the count required for retention, so that $\hat S=\{j:N_j\ge r\}$.
To describe retention within a feature set $G$, we distinguish distinct features from selection occurrences.
A selection occurrence is one feature selected in one split; a feature selected in several splits contributes several occurrences.

Write $A_G$ for the mean number of selections from $G$ per split and $I_G$ for the mean pairwise overlap within $G$, including self-pairs $(m=m')$:
\begin{equation}
 A_G=\sum_{j\in G}\Pi_j,\qquad
 I_G=\sum_{j\in G}\Pi_j^2.
 \label{eq:recurrence-definitions}
\end{equation}
When $A_G>0$, the recurrence ratio $\rho_G$ normalizes this overlap by the mean selection count, while $F_G$ measures the fraction of selection occurrences attributable to features retained in the final support:
\begin{equation}
 \rho_G=I_G/A_G,\qquad
 F_G=\frac{1}{A_G}\sum_{j\in G\cap\hat S}\Pi_j.
 \label{eq:recurrence-ratios}
\end{equation}
These quantities describe the finite collection of $M$ recorded split selections.
When $G$ contains all features, $\rho_G$ is closely related to the stability index of \citet{nogueira2018stability}.

\begin{proposition}[Recurrence bounds]
\label{prop:recurrence}
For any $M$ split selections, any $\eta\in(0,1)$, and any $G$ with $A_G>0$,
\begin{equation}
 \frac{\rho_G-\eta}{1-\eta}\le F_G\le\frac{\rho_G}{\eta},
 \qquad
 \frac{I_G-\eta A_G}{1-\eta}\le |\hat S\cap G|
 \le\frac{I_G}{\eta^2}.
 \label{eq:recurrence-sandwich}
\end{equation}
For $\eta=1$, the upper bounds remain valid.
\end{proposition}

The first pair of bounds describes the fraction of selection occurrences retained; the second describes the number of distinct features retained.
When $\rho_G>\eta$, the lower bounds ensure that some features survive aggregation, while the upper bounds quantify how limited agreement restricts retention.
If the true support $S^\star$ is known, setting $G=S^\star$ gives the power bound $|\hat S\cap S^\star|/|S^\star|\le I_{S^\star}/(\eta^2|S^\star|)$.
This provides a retention diagnostic for known-support benchmarks using the same recorded selections that form the aggregate.

\paragraph{False selections under repeated randomization.}
We now condition on the observed dataset $D$ and fix the procedure, including its preprocessing, feature mapping, and failure policy.
Assume that all split-specific random inputs are independent and identically distributed (IID) conditional on $D$, and that every attempted repetition returns a measurable selected set under the fixed failure policy.
Overlap between row partitions is compatible with this model because the dataset is held fixed; the independence assumption concerns the random inputs to each repetition.
The conditional-binomial calculation follows \citet[Section~3.3]{ren2023derandomizing}.
For STK, the random inputs govern the complete screening and testing procedure, rather than only repeated knockoff draws on a fixed tested feature family.

Let $\theta_j(D)=\Pr(j\in\hat S^{(1)}\mid D)$ be the probability that one repetition selects feature $j$.
Fix a nonempty scientific null set $H_0$, and let $V=|\hat S\cap H_0|$ count false selections in the aggregate.
The count $r=\lceil M\eta\rceil$ is the minimum number of selections required for retention.
For a feature selected with probability $u$ in each repetition, $B_{M,r}(u)$ is the probability of being selected at least $r$ times across $M$ repetitions.
\begin{equation}
 B_{M,r}(u)=\Pr\{\mathrm{Bin}(M,u)\ge r\},
 \qquad \theta_{\max}(D)=\max_{j\in H_0}\theta_j(D).
 \label{eq:binomial-tail}
\end{equation}
Under the conditional IID model, the probability that aggregation retains feature $j$ is $B_{M,r}(\theta_j(D))$, and the expected false-selection count is $\mathbb{E}[V\mid D]=\sum_{j\in H_0}B_{M,r}(\theta_j(D))$.
The following bound compares this count with that of one repetition.

\begin{corollary}[Diffuse-null count bound]
\label{cor:null-count}
Let $V_1=|\hat S^{(1)}\cap H_0|$ and $\kappa_{M,r}(t)=\sup_{0<u\le t}B_{M,r}(u)/u$.
If $\theta_{\max}(D)>0$, then
\begin{equation}
 \mathbb{E}[V\mid D]\le
 \kappa_{M,r}\!\bigl(\theta_{\max}(D)\bigr)\,
 \mathbb{E}[V_1\mid D]
 \le \min_{1\le s\le r}\frac{\binom{M}{s}}{\binom{r}{s}}
 \theta_{\max}(D)^{s-1}\mathbb{E}[V_1\mid D].
 \label{eq:null-count-bound}
\end{equation}
If $\theta_{\max}(D)=0$, then $V=0$ almost surely.
\end{corollary}

For a feature selected with probability $u$ in one repetition, the ratio $B_{M,r}(u)/u$ compares its selection probability after aggregation with that before aggregation.
The factor $\kappa_{M,r}(\theta_{\max}(D))$ takes the worst-case ratio over all probabilities up to the largest null-feature selection probability.
A factor below one establishes a reduction in the expected false-selection count under the conditional model; a factor above one does not establish a reduction.
The rightmost expression gives an explicit upper bound on this multiplier, with the minimum over $s$ choosing the tightest of the available bounds.

At the evaluated setting $M=25$ and $\eta=0.5$, a feature must be selected in at least $r=13$ repetitions.
Under the conditional model, this majority rule lowers inclusion probabilities between zero and one half and raises those between one half and one.
Aggregation can therefore suppress infrequent false selections while reinforcing persistent ones; the same rule can remove signals that are selected too infrequently.

The conditional probabilities, including $\theta_{\max}(D)$, are unknown, and agreement among recorded selections does not establish whether the recurring features are signals.
These results characterize retention but do not provide aggregate FDR or PFER control.
Appendix~\ref{app:theory} gives the proofs, finite-repetition qualifications, and examples.
The matched analysis in Section~\ref{sec:matched} measures the actual changes in FDP and power when the same split outputs are aggregated.

Joint-selection bounds are used in stability-selection theory to relate repeated selection to retention \citep{shah2013variable}.
Here we use a related counting argument to bound retention for a fixed collection of split selections.

%% file: tables/matched_compact.tex
\begin{tabular}{@{}l r l l@{}}
\toprule
Method & $k$ & $\Delta$FDP & $\Delta$power \\
\midrule
Headline STK & 3 & -0.230 [-0.283, -0.176] & +0.004 [-0.002, +0.010] \\
Headline STK & 10 & -0.209 [-0.245, -0.174] & +0.023 [-0.029, +0.074] \\
Headline STK & 20 & -0.255 [-0.327, -0.184] & -0.037 [-0.062, -0.012] \\
Secondary STK & 3 & -0.275 [-0.330, -0.221] & +0.008 [+0.001, +0.015] \\
Secondary STK & 10 & -0.194 [-0.242, -0.147] & +0.040 [-0.008, +0.088] \\
Secondary STK & 20 & -0.213 [-0.277, -0.150] & +0.014 [-0.010, +0.037] \\
\bottomrule
\end{tabular}

%% file: appendix.tex
\input{theory_appendix}

\section{Additional empirical comparisons}

\subsection{DepMap comparison with pooled STK}
\label{app:depmap-pooled}

Table~\ref{tab:depmap-pooled} supplements the main DepMap results with the pooled STK reference.
Pooled STK screens and tests candidates on the same observations, without cross-split aggregation.
Its evaluated knockoff+ level is 0.1 with offset 1, so it returns at least ten features or none; exact recovery at $k=3$ is therefore impossible under this threshold.
At $k=3$, it recovered all three support features among ten selections in eight datasets and abstained in two; at $k=10$, it had higher mean power but higher mean FDP than headline STK.
These contrasts are descriptive, not a clean aggregation ablation: the configurations differ in sample use, aggregation, per-filter level, and fitting budget.

\begin{table}[htbp]
\centering
\caption{Pooled STK on the semi-synthetic DepMap benchmark.
Values are means over ten datasets per true support size; selected size counts all retained features.}
\label{tab:depmap-pooled}
\input{tables/depmap_pooled}
\end{table}

\subsection{Additional matched aggregation results}
\label{app:matched}

Table~\ref{tab:matched-additional} reports the support sizes omitted from the main matched table.
The study used the Gaussian density-sweep design, ten datasets per signal support size (seeds 42--51), and 25 splits per configuration and dataset.
Individual-split metrics were averaged within each dataset before comparison with the aggregate of those same outputs; only aggregation changes in this contrast.
Of 140 signal aggregates, 137 matched the supports reported in the Gaussian density sweep.
Every paired calculation therefore uses the matched study's own records, not substituted density-sweep aggregates.

\begin{table}[htbp]
\centering
\caption{Additional matched aggregate minus mean individual-split results.
Brackets are descriptive, unadjusted 95\% $t$ intervals across ten dataset-level differences, not 250 independent splits.
Together with Table~\ref{tab:matched-compact}, these rows cover all evaluated support sizes $k\geq2$.}
\label{tab:matched-additional}
\input{tables/matched_additional}
\end{table}

\subsection{PC-Knockoff comparisons}
\label{app:pc}
\label{sec:pc}

The PC comparison used the Gaussian design in Table~\ref{tab:benchmarks}, with signal seeds 10000--10009 per evaluated support size.
Global-null datasets used seeds 10000--10029.
All methods used the same feature matrices and outcomes within each dataset; these datasets are separate from the density-sweep and matched studies.
PC was evaluated at $\alpha\in\{0.1,0.2,0.5\}$ and four allocations $(n_{\rm screen},d)$, with $800-n_{\rm screen}$ testing rows.
Repeated PC reran screening and testing over 25 splits and retained features selected in at least 13 splits.
The allocations and original compact comparison rule were fixed before signal results were seen; STK aggregates were reused from the shared-dataset comparison.

\begin{table}[htbp]
\centering
\caption{Mean power / mean FDP for STK and PC at PC threshold $\alpha=0.5$.
PC row labels give $(n_{\rm screen},d)$.
The mean-split row averages metrics over 25 individual splits within each dataset; aggregate rows apply frequency thresholding to the split outputs.
Values are means over ten datasets per support size, not split-0 results.}
\label{tab:pc-compact}
\input{tables/pc_appendix}
\end{table}

At the original 200/50 allocation, aggregation reduced PC's mean FDP but also its mean power at all three displayed support sizes; stricter PC thresholds often abstained.
Larger screening allocations narrowed the power gap with headline STK.
For repeated PC at 400/50 and $\alpha=0.5$, headline STK minus PC power differences were $0.030\;[-0.077,0.137]$ at $k=10$ and $-0.010\;[-0.060,0.040]$ at $k=20$, using descriptive, unadjusted 95\% paired $t$ intervals across datasets.
Across all four allocations and three levels, repeated PC selected features in at most 1 of 30 global-null datasets per setting; both STK configurations selected none on these null datasets.

The PC-Knockoff implementation follows the authors' code \citep{liu2022modelfree}: projection-correlation screening, fixed-X equicorrelated construction from the held-out sample Gram matrix, and a projection-correlation squared-difference statistic.
STK instead uses a multivariate stability screen, estimated-Gaussian knockoffs, and lasso coefficient differences.
PC's exactness and sure screening are not certified here, so its published conditional guarantee is not asserted for these benchmarks or the repeated adaptation.
The comparisons neither isolate the screen alone nor impose equal computational cost or a common FDR guarantee.
Single-split PC had a pooled median runtime of 4.06~s on the same HPC evaluation, compared with 77.35~s for headline STK and 123.73~s for secondary STK.
Repeated-PC timings came from a laptop with concurrent processes and are not directly comparable with these HPC times.

\paragraph{Reproducibility.}
The code and result package should be used with each study's frozen configurations, software versions, and source hashes, rather than current package defaults.
Both STK configurations were fixed before the matched and PC studies; the frequency threshold followed synthetic development, not prospective error calibration.
Screening-row preprocessing removes constants and scales without centering; its transformation is applied to testing rows, where candidates are rescaled.
Screening also includes a pooled knockoff fit whose selections are discarded and whose cost is included in end-to-end timing.
Permutation fallback does not generally satisfy Model-X exchangeability.
Held-out STK runs in the matched and PC studies recorded no fallback; Gaussian density-sweep and DepMap runs did not retain fallback status.
Recorded solver warnings do not establish numerical insensitivity.
Detailed operating-point tables, paired intervals, and recurrence diagnostics are retained with the result materials rather than repeated here.

\section{Supporting global-null and scaling results}

\begin{figure}[htbp]
\centering
\includegraphics[width=\textwidth]{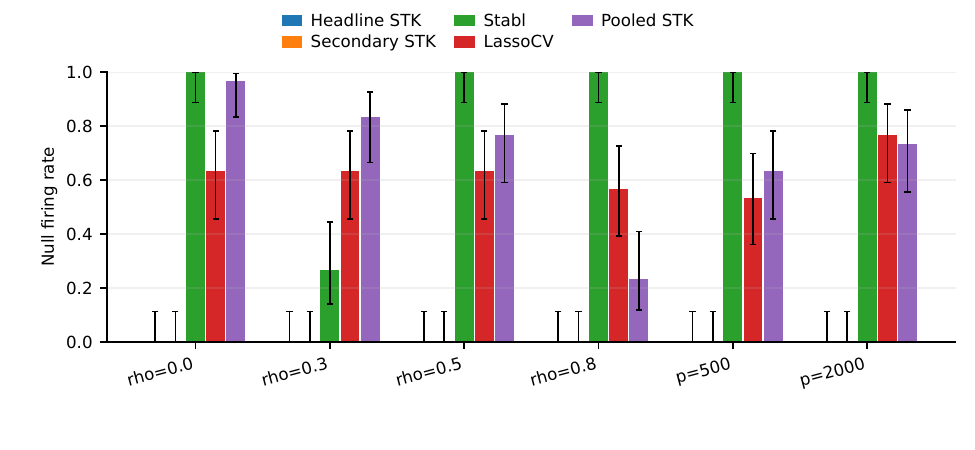}
\caption{\textbf{False selections under the Gaussian global null.}
Probability of selecting any feature across correlation and dimension sweeps, with $n=800$ and targets independent of the features; error bars show Wilson 95\% confidence intervals over 30 datasets per setting.
Correlation varies at $p=5000$, and dimension varies at $\rho=0.5$.
Both multi-split STK configurations made no selections across the 180 evaluated datasets.
These datasets are separate from the matched and PC null studies; their counts are not pooled.}
\label{fig:null}
\end{figure}

\begin{figure}[htbp]
\centering
\includegraphics[width=\textwidth]{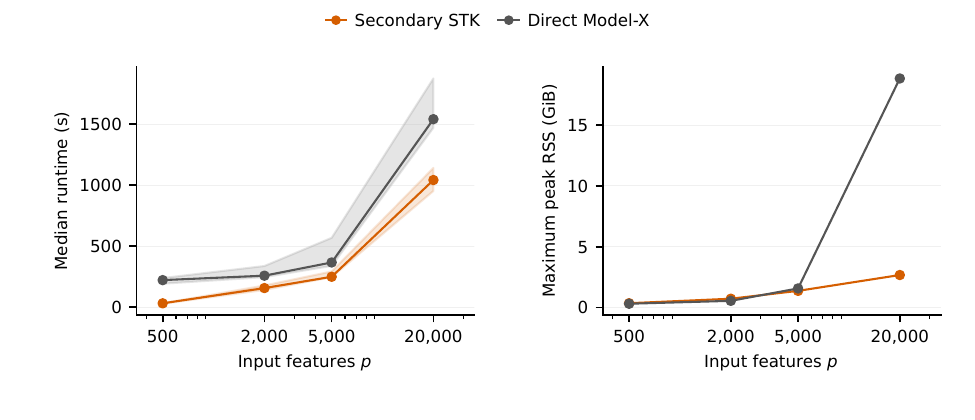}
\caption{\textbf{Computational scaling.}
Secondary STK and direct full-dimensional Model-X were evaluated on ten Gaussian datasets per dimension, with $n=800$, $k=3$, $\rho=0.8$, signal-to-noise ratio 0.5, and 16 CPUs and 64~GB per run.
Runtime lines show medians with interquartile bands, rather than the means reported in the main text; memory is the maximum resident set over the ten datasets.
Secondary STK uses a 70/30 screening/testing allocation, candidate floor $K=20$, and 25 splits.}
\label{fig:tractability}
\end{figure}

%% file: theory_appendix.tex
\section{Supporting theory}
\label{app:theory}

\subsection{Deterministic recurrence bounds}

Write $N_j=M\Pi_j$, $r=\lceil M\eta\rceil$, and $\hat S=\{j:N_j\ge r\}$.
For a fixed $G$ with $A_G>0$, let $R=G\cap\hat S$ and $m_R=\sum_{j\in R}\Pi_j=F_GA_G$.
Exchanging sums gives
\[
 A_G=\frac1M\sum_m|\hat S^{(m)}\cap G|,
 \qquad
 I_G=\frac1{M^2}\sum_{m,m'}
 |\hat S^{(m)}\cap\hat S^{(m')}\cap G|.
\]
The second identity includes self-pairs.
If one draws a selection occurrence uniformly and then a split uniformly, $\rho_G=I_G/A_G$ is the chance that the selected feature recurs in that split.

\begin{proof}[Proof of Proposition~\ref{prop:recurrence}]
For $j\in R$, $\Pi_j\ge\eta$, hence $\Pi_j^2\ge\eta\Pi_j$.
Thus $I_G\ge\eta m_R$, giving $F_G\le\rho_G/\eta$ and $|R|\le m_R/\eta\le I_G/\eta^2$.
For the lower bounds, $\Pi_j^2\le\Pi_j$ on $R$, while $\Pi_j^2\le\eta\Pi_j$ outside $R$.
Consequently,
\[
 I_G\le m_R+\eta(A_G-m_R),
\]
which gives $m_R\ge(I_G-\eta A_G)/(1-\eta)$ for $0<\eta<1$.
Divide by $A_G$ for the bound on $F_G$, and use $|R|\ge m_R$ for the count bound.
The upper bounds use only $\eta>0$ and therefore also hold at $\eta=1$.
\end{proof}

For the full set of $p$ features, with $0<A<p$ and $M>1$, the stability estimator $\hat\Phi$ of \citet{nogueira2018stability} uses indicator variances $s_j^2=\{M/(M-1)\}\Pi_j(1-\Pi_j)$.
Since $\sum_j\Pi_j(1-\Pi_j)=A-I$, its relation to recurrence is
\[
 \rho=1-\frac{M-1}{M}(1-A/p)(1-\hat\Phi).
\]
Thus recurrence is related to an established stability measure, not a proposed new index.

\subsection{Conditional randomization and finite repetitions}

Fix $D=(X,y)$ in the original feature space and write $\hat S^{(m)}=g(D,U_m)$, where $U_m$ contains all split-specific randomness.
Assume that the procedure, preprocessing, feature mapping, and failure policy are fixed; that the $U_m$ are IID conditional on $D$; and that every attempted split returns a measurable set.
A prespecified failed split may return the empty set and still count in the denominator; dropping failed splits changes the procedure.
Overlapping row partitions are compatible with this model because $D$ is fixed, but are not independent datasets.
These assumptions are idealized: pseudorandom seed schedules do not prove IID randomization, and the software can abort on uncaught errors.
The analysis covers the original-space frequency tally before any subsequent filtering by the estimator's global feature mask.

Set $\theta_j(D)=\Pr\{j\in g(D,U_1)\mid D\}$ and fix a nonempty scientific null set $H_0$.
With $B_{M,r}$ as in Equation~\eqref{eq:binomial-tail}, conditional IID repetitions give
\begin{equation}
 \Pr(j\in\hat S\mid D)=B_{M,r}(\theta_j(D)),\qquad
 \mathbb{E}[|\hat S\cap H_0|\mid D]=
 \sum_{j\in H_0}B_{M,r}(\theta_j(D)).
 \label{eq:app-binomial}
\end{equation}
No independence between features is needed for the second identity.

\begin{proof}[Proof of Corollary~\ref{cor:null-count}]
Equation~\eqref{eq:app-binomial} gives $\mathbb{E}[V\mid D]=\sum_{j\in H_0}B_{M,r}(\theta_j)$.
For $\theta_j>0$, its summand is at most $\kappa_{M,r}(\theta_{\max})\theta_j$; summing gives the first bound.
For $s\in\{1,\ldots,r\}$, a feature with count $N_j$ belongs to exactly $\binom{N_j}{s}$ intersections of $s$ split selections, at least $\binom{r}{s}$ if retained.
Summing over any fixed $G$ gives the pathwise counting inequality
\begin{equation}
 \binom{r}{s}|\hat S\cap G|\le
 \sum_{T\subseteq\{1,\ldots,M\}:|T|=s}
 \left|G\cap\bigcap_{m\in T}\hat S^{(m)}\right|.
 \label{eq:app-counting}
\end{equation}
For $N\sim\mathrm{Bin}(M,u)$, the same counting argument and $\mathbb{E}\binom Ns=\binom Ms u^s$ give
\[
 B_{M,r}(u)\le
 \frac{\binom{M}{s}u^s}{\binom{r}{s}}.
\]
Divide by $u$, take the supremum over $0<u\le\theta_{\max}$, and minimize over $s$ for the explicit multiplier bound.
If $\theta_{\max}=0$, every summand is zero.
\end{proof}

Define the ideal thresholded probability set $S_\eta(D)=\{j:\theta_j(D)\ge\eta\}$.
If every feature satisfies $|\theta_j(D)-\eta|\ge\Delta>0$, Hoeffding's inequality \citep{hoeffding1963} and a union bound give
\begin{equation}
 \Pr\{\hat S\ne S_\eta(D)\mid D\}
 \le \min\{1,p\exp(-2M\Delta^2)\}.
 \label{eq:app-concentration}
\end{equation}
This is concentration around an algorithmic probability set on one fixed dataset; it neither identifies $S_\eta(D)$ with a true support nor selects $\eta$.
At $\theta_j(D)=\eta$, the positive-margin condition fails and a limiting inclusion decision need not converge.
In particular, with $\eta=1/2$ and odd $M$, binomial symmetry gives inclusion probability one half at the boundary.

Conditional on $D$, two distinct IID splits have expected overlap $\sum_{j\in G}\theta_j(D)^2$, whereas the finite statistic includes self-pairs:
\[
 \mathbb{E}[\Pi_j^2\mid D]=\theta_j(D)^2+
 \frac{\theta_j(D)[1-\theta_j(D)]}{M}.
\]
For $M>1$, the off-diagonal overlap $(M(M-1))^{-1}\sum_{m\ne m'}|G\cap\hat S^{(m)}\cap\hat S^{(m')}|$ is unbiased for the conditional population overlap, but its ratio with an estimate of $A_G$ is not generally unbiased.
It does not replace $I_G$ in the deterministic finite-split bounds.

\subsection{Why the count result is not aggregate error control}

The pathwise $s=1$ case gives $rV\le\sum_m|\hat S^{(m)}\cap H_0|$.
Even a per-split PFER bound is therefore multiplied by $M/r$, which exceeds one at $M=25,r=13$.
FDR has a further random denominator and does not follow from this count bound.

For an explicit IID selector counterexample, let $z$ be null and $t_1,\ldots,t_{25}$ be non-null.
Independently draw $J_m$ uniformly from $\{1,\ldots,25\}$ and return $\hat S^{(m)}=\{z,t_{J_m}\}$, giving FDP $1/2$ in every split.
At $M=25$, $\eta=1/2$, $z$ always enters the aggregate and at most one non-null feature can reach 13 selections, so aggregate FDP is at least $1/2$.
On the positive-probability event that all $J_m$ are distinct, aggregate FDP is one; hence aggregate FDR exceeds $1/2$.
As $M$ grows, the non-null frequencies approach $1/25$ and aggregate FDP approaches one.
This generic selector is not a construction of valid Model-X outputs and does not show that STK has this behavior; it shows only that per-split FDR alone does not transfer to a frequency aggregate.

Screening can also change the tested hypothesis: if $Y=X_1+\varepsilon$ and $X_2$ is correlated with $X_1$, then $X_2$ can be null conditional on all other features yet associated with $Y$ after a screen retains $X_2$ but omits $X_1$.
Sample separation does not remove this change from a full-model to a reduced-family target.
These limits, together with approximate knockoffs and any permutation fallback, prevent an aggregate FDR or PFER claim for the present analysis.

%% file: tables/depmap_pooled.tex
\begin{tabular}{@{}rrrr@{}}
\toprule
$k$ & Mean power & Mean FDP & Mean selected size \\
\midrule
3 & 0.800 & 0.560 & 8.0 \\
10 & 0.830 & 0.232 & 10.8 \\
\bottomrule
\end{tabular}

%% file: tables/matched_additional.tex
\begin{tabular}{@{}lrll@{}}
\toprule
Method & $k$ & $\Delta$FDP & $\Delta$power \\
\midrule
Headline STK & 2 & -0.257 [-0.336, -0.179] & +0.004 [-0.005, +0.013] \\
Headline STK & 5 & -0.232 [-0.263, -0.201] & +0.043 [+0.008, +0.079] \\
Headline STK & 15 & -0.230 [-0.306, -0.154] & -0.037 [-0.079, +0.006] \\
Secondary STK & 2 & -0.267 [-0.328, -0.206] & +0.018 [+0.001, +0.035] \\
Secondary STK & 5 & -0.254 [-0.309, -0.198] & +0.022 [+0.003, +0.042] \\
Secondary STK & 15 & -0.201 [-0.238, -0.165] & +0.021 [-0.018, +0.059] \\
\bottomrule
\end{tabular}

%% file: tables/pc_appendix.tex
\begin{tabular}{@{}lccc@{}}
\toprule
Method & $k=3$ & $k=10$ & $k=20$ \\
\midrule
Headline STK & 1.000 / 0.075 & 0.570 / 0.000 & 0.150 / 0.000 \\
Secondary STK & 1.000 / 0.115 & 0.830 / 0.050 & 0.310 / 0.064 \\
PC mean split (200/50) & 0.843 / 0.457 & 0.311 / 0.548 & 0.105 / 0.569 \\
PC aggregate (200/50) & 0.833 / 0.108 & 0.200 / 0.040 & 0.030 / 0.000 \\
PC aggregate (200/100) & 0.900 / 0.147 & 0.370 / 0.054 & 0.050 / 0.040 \\
PC aggregate (400/50) & 0.967 / 0.222 & 0.540 / 0.119 & 0.160 / 0.115 \\
PC aggregate (400/100) & 0.900 / 0.065 & 0.500 / 0.125 & 0.130 / 0.058 \\
\bottomrule
\end{tabular}

%% file: references.bib
@article{poste2011biomarkers,
  title={Bring on the biomarkers},
  author={Poste, George},
  journal={Nature},
  volume={469}, number={7329}, pages={156--157}, year={2011},
  doi={10.1038/469156a}
}

@article{eindor2005outcome,
  title={Outcome signature genes in breast cancer: Is there a unique set?},
  author={Ein-Dor, Liat and Kela, Itai and Getz, Gad and Givol, David and Domany, Eytan},
  journal={Bioinformatics},
  volume={21}, number={2}, pages={171--178}, year={2005},
  doi={10.1093/bioinformatics/bth469}
}

@article{wasserman2009highdimensional,
  title={High-dimensional variable selection},
  author={Wasserman, Larry and Roeder, Kathryn},
  journal={The Annals of Statistics},
  volume={37}, number={5A}, pages={2178--2201}, year={2009},
  doi={10.1214/08-AOS646}
}

@article{ambroise2002selection,
  title={Selection bias in gene extraction on the basis of microarray gene-expression data},
  author={Ambroise, Christophe and McLachlan, Geoffrey J},
  journal={Proceedings of the National Academy of Sciences},
  volume={99}, number={10}, pages={6562--6566}, year={2002},
  doi={10.1073/pnas.102102699}
}

@article{tibshirani1996regression,
  title={Regression shrinkage and selection via the lasso},
  author={Tibshirani, Robert},
  journal={Journal of the Royal Statistical Society: Series B},
  volume={58}, number={1}, pages={267--288}, year={1996}
}

@article{zou2005regularization,
  title={Regularization and variable selection via the elastic net},
  author={Zou, Hui and Hastie, Trevor},
  journal={Journal of the Royal Statistical Society: Series B},
  volume={67}, number={2}, pages={301--320}, year={2005}
}

@article{meinshausen2010stability,
  title={Stability selection},
  author={Meinshausen, Nicolai and B{\"u}hlmann, Peter},
  journal={Journal of the Royal Statistical Society: Series B},
  volume={72}, number={4}, pages={417--473}, year={2010}
}

@article{barber2015controlling,
  title={Controlling the false discovery rate via knockoffs},
  author={Barber, Rina Foygel and Cand{\`e}s, Emmanuel J},
  journal={The Annals of Statistics},
  volume={43}, number={5}, pages={2055--2085}, year={2015}
}

@article{barber2019highdimensional,
  title={A knockoff filter for high-dimensional selective inference},
  author={Barber, Rina Foygel and Cand{\`e}s, Emmanuel J},
  journal={The Annals of Statistics},
  volume={47}, number={5}, pages={2504--2537}, year={2019},
  doi={10.1214/18-AOS1755}
}

@article{candes2018panning,
  title={Panning for gold: Model-X knockoffs for high dimensional controlled variable selection},
  author={Cand{\`e}s, Emmanuel and Fan, Yingying and Janson, Lucas and Lv, Jinchi},
  journal={Journal of the Royal Statistical Society: Series B},
  volume={80}, number={3}, pages={551--577}, year={2018}
}

@article{ren2023derandomizing,
  title={Derandomizing knockoffs},
  author={Ren, Zhimei and Wei, Yuting and Cand{\`e}s, Emmanuel},
  journal={Journal of the American Statistical Association},
  volume={118}, number={542}, pages={948--958}, year={2023}
}

@article{hedou2024stabl,
  title={Discovery of sparse, reliable omic biomarkers with {Stabl}},
  author={H{\'e}dou, Julien and others},
  journal={Nature Biotechnology}, volume={42}, number={10}, pages={1581--1593}, year={2024},
  doi={10.1038/s41587-023-02033-x}
}

@article{liu2022modelfree,
  title={Model-free feature screening and {FDR} control with knockoff features},
  author={Liu, Wanjun and Ke, Yuan and Liu, Jingyuan and Li, Runze},
  journal={Journal of the American Statistical Association},
  volume={117}, number={537}, pages={428--443}, year={2022},
  doi={10.1080/01621459.2020.1783274}
}

@misc{depmap2026q1,
  author={{Broad Institute DepMap}},
  title={{DepMap 26Q1 Public}},
  year={2026},
  howpublished={DepMap Portal},
  url={https://forum.depmap.org/t/announcing-the-26q1-release/4606},
  note={Accessed 2026-08-15}
}

@article{hoeffding1963,
  author = {Hoeffding, W.},
  title = {Probability Inequalities for Sums of Bounded Random Variables},
  journal = {Journal of the American Statistical Association},
  volume = {58},
  number = {301},
  pages = {13--30},
  year = {1963},
  doi = {10.1080/01621459.1963.10500830}
}

@article{shah2013variable,
  author  = {Shah, Rajen D. and Samworth, Richard J.},
  title   = {Variable Selection with Error Control: Another Look at Stability Selection},
  journal = {Journal of the Royal Statistical Society: Series B (Statistical Methodology)},
  volume  = {75},
  number  = {1},
  pages   = {55--80},
  year    = {2013},
  doi     = {10.1111/j.1467-9868.2011.01034.x}
}

@article{nogueira2018stability,
  author  = {Nogueira, Sarah and Sechidis, Konstantinos and Brown, Gavin},
  title   = {On the Stability of Feature Selection Algorithms},
  journal = {Journal of Machine Learning Research},
  volume  = {18},
  number  = {174},
  pages   = {1--54},
  year    = {2018}
}
